\documentstyle[aps,amsfonts,amssymb,multicol]{revtex}
\input{psfig}

\def\veps{\varepsilon}

\def\be{\begin{equation}}
\def\ee{\end{equation}}
\def\bea{\begin{eqnarray}}
\def\eea{\end{eqnarray}}

\begin{document}
\draft
\def\vareps{\varepsilon}
\title{Correlation--enhanced Friedel oscillations in amorphous alloys
and quasicrystals}
\author{Johann Kroha$^{\dag}$}
\address
{Universit\"at Karlsruhe, Institut f\"ur Theorie der Kondensierten
Materie, Postfach 6980, 76128 Karlsruhe, Germany}
\date{\today}
\maketitle
\begin{abstract}                                           
We show that quantum correlations induced
by electron--electron interactions in the presence of random impurity
scattering can play an important role in the thermal
stabilization of amorphous Hume--Rothery systems: When 
there is strong backscattering off local, concentrical
ion clusters, the static electron
density response $\chi (0,q)$ acquires a powerlaw divergence at $q=2k_F$
even at elevated temperature. 
This leads to an enhancement as well as
to a systematical phase shift of the Friedel oscillations, 
both consistent with experiments.
The possible importance of this effect in icosahedral quasicrystals 
is discussed.
\end{abstract}
\begin{multicols}{2}
\section{introduction}
A large class of noble--polyvalent metal alloys exhibit a crystalline to 
amorphous transformation (CAT) as a function of the 
polyvalent metal content. There are several experimental indications
for the amorphous phase being stabilized by the Hume--Rothery (HR)
mechanism, i.e. by forming a structure--induced pseudogap in the 
electronic density of states (DOS) at the Fermi level $\veps _F$. 
The experimental evidence \cite{haeuss.92} includes 
the observation of a pronounced pseudogap, 
a maximum of the electrical resistivity at the CAT, and the coincidence of
the atomic spacing $a$ with the Friedel wave length $\lambda _F = \pi /k_F$
near the CAT, where $\hbar k_F$ is the Fermi momentum.
Very similar behavior is found in icosahedral ({\it i})
quasicrystals \cite{davydov.96,poon.92}. 
The conjecture of a HR--like stabilization mechanism is strongly supported by 
detailed theoretical studies both for amorphous \cite{hafner.90}
and for quasicrystalline 
\cite{ashcroft.87,fuji.91,hafner.92,dmitrienko.95} systems,
although there is also the
possibility of entropic stabilization \cite{joseph.97}. 
The fact that in the amorphous state 
the above--mentioned structural matching 
is observed over distances of up to $5\lambda _F$ has lead to the
assumption that these remarkably long--range correlations are induced by
the ions being bound in the minima of the potential formed by the
Friedel oscillations (FO) around an arbitrary central ion.
 
However, several experimental puzzles have remained unexplained:
(1) At finite temperature $T$ and also in the presence of disorder the 
impurity--averaged FOs are exponentially damped 
due to the spread of the Fermi momentum over a width given by $T$ 
and the inverse elastic mean free path, respectively. Hence,
the stability of amorphous alloys
at elevated $T$, in particular their intermediate--range structural 
correlations, are difficult to explain by the {\it conventional} FOs.
(2) In all HR--systems, invariably the amorphous state is thermally 
most stable just at the CAT \cite{haeuss.92}.
(3) In the amorphous state the ionic positions are systematically 
shifted compared to the minima of the Friedel potential of a free 
electron sea \cite{haeuss.92}. The experimental
findings (1)--(3) raise the question of a systematical, composition
dependent enhancement and phase shift $\varphi$ of the FOs,
where $\varphi$ varies from $\varphi \simeq \pi $ deep inside the amorphous 
phase to  $\varphi = \pi /2$ at the CAT. As shown below,
these problems can be explained by one single quantum effect. 
 
\section{Theory}
In amorphous alloys the electronic motion is diffusive instead of ballistic.
Since diffusion, as a dissipative process, is difficult to include in an
{\it ab initio} calculation, we here 
choose a Feynman diagram technique, where diffusion arises
in the formalism by averaging over all random configurations 
of the system. 
The Nagel--Tauc condition $2k_F \simeq  2\pi /a \equiv k_p$
implies strong electronic backscattering off local ion clusters. 
It has been shown \cite{mahan.90,kroha.90,kroha.95} that this 
not only generates a pseudogap 
but at the same time leads to a substantial enhancement of the 
electron transport or density relaxation rate $\tau ^{-1}$ over the 
quasiparticle decay rate $\tau _{qp}^{-1}$. This is evidenced 
experimentally by the anomalously small electrical 
conductivity, $\sigma = ne^2\tau /m^{*}$. 
Thus, we have as a generic feature of the amorphous state:
$\tau ^{-1} \gg \tau _{qp}^{-1}$.

\begin{figure} 
\centerline{\psfig{figure=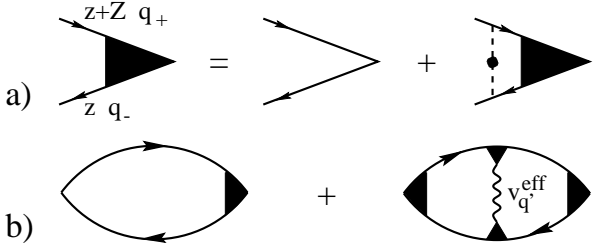,rheight=4.2cm}}
\narrowtext
\caption{
a) Diffusion vertex $\Gamma$. b) Polarisation $\Pi (0,q)$
including leading order quantum correction induced by disorder and 
interactions. Dashed lines denote electron--ion scattering, 
the wavy line with solid triangles the effective Coulomb interaction.
\label{fig:polaris}}
\end{figure}

We now turn to the calculation of the electron
density distribution $\rho (r)$ around an ion embedded in the 
disordered electron sea. 
One does not expect a single ion to generate a sizable phase shift
of the FOs at distances $r>a$, $\rho (r) \propto 
\mbox{cos}(2k_Fr-\varphi )/r^{3}$,  
since it is equal to the electron backscattering phase off
that ion, which is small unless $\veps _F$ is close 
to an internal resonance. Instead we consider quantum effects due to
disorder and Coulomb interaction
and assume, for simplicity, a point--like ion charge and a quadratic
band $\veps _k = \hbar ^2 k^2/2m^{*}$. 
In a diffusive electron system screening is inhibited, so that the
effective Coulomb interaction $v_q^{eff} (z,Z)$ between electrons with  
complex frequencies $z$ and $z+Z$ acquires a long--range,   
retarded part \cite{altsh.79},
\begin{eqnarray}
v_q^{eff}(z,Z)={v_q \over \epsilon ^{RPA}(Z,q)} \Gamma ^2(z,Z,q),\qquad
v_q={4\pi e^2 \over q^2},                 
\label{veff}
\end{eqnarray}
where $\epsilon ^{RPA}(Z,q)\! =\! 1\! +\! 2\pi i\ \sigma\ /
(Z\, \mbox{sgn}Z''+iq^2D)$ is the disordered 
RPA dynamical dielectric function and the diffusion vertex, defined in
Fig.~\ref{fig:polaris} a), is
\begin{eqnarray}
\Gamma (z,Z,q) = \left\{  \begin{array}{ll}
       {i/\tau  \ \mbox{sgn}Z'' \over  
       Z+iq^2D\ \mbox{sgn}Z'' }\quad &z''(z+Z)''<0  \\
       1 & \mbox{otherwise.}
                          \end{array}
                 \right.                   
\label{vertex}
\end{eqnarray}
$D=1/3\  v_F^2\tau $  
and $''$ denote the diffusion constant
and 
\noindent
the imaginary part, respectively. 
The long--range nature of $v_q^{eff}$ originates from the 
diffusion pole of $\Gamma$.
In order to calculate its effect on the FOs, 
one must consider contributions to the polarisation $\Pi (0,q)$ where 
$\Gamma (z,Z,q')$ enters at $Z,q' \simeq 0$, although $\Pi (0,q)$ is 
evaluated at large external momenta $q\simeq 2k_F$.  
The leading singular contribution  
arises from the quantum correction shown in the second diagram
of Fig.~\ref{fig:polaris} b). For amorphous metals 
($\tau^{-1}\gg \tau_{qp}^{-1}$, see above) 
it is evaluated as \cite{kroha.95},
\begin{eqnarray}
\Pi ^{(1)}(0,q) = C(\varepsilon _F\tau)
\int _{-\veps_F}^{\veps_F} \mbox{d}\nu 
{1/(4T) \over \mbox{cosh}^2{\nu\over 2T}} 
{\mbox{sgn}(x-1)\over \sqrt{|x-1|}},
\label{Pi3}
\end{eqnarray}
where $x=x(\nu )=(q/ 2k_F) / \sqrt{1+\nu / \veps _F}$ and 
$C(\veps _F\tau )=-0.343~[2m^{*}k_F/(2\pi \hbar )^2]/
(\veps _F\tau )^{7/2}$.
The first term of Fig.~\ref{fig:polaris} b), $\Pi ^{(0)}(0,q)$,
corresponds to the Lindhard function, where 
$\Gamma $ contributes only a nonsingular factor of ${\cal O}(1)$. 
It is seen from Eq.~(\ref{Pi3}) that for $T=0$, 
$\Pi ^{(1)}(0,q)$ exhibits a powerlaw divergence 
$\propto -\mbox{sgn}(q-2k_F)/|q-2k_F|^{1/2}$ 
at $q=2k_F$.
Although at finite $T$ or $\tau _{qp}^{-1}$ the divergence 
of $\Pi ^{(1)}(0,q)$ is reduced to a peak, the inverse dielectric 
function $1/\veps (q) =1/(1-v_q \Pi (0,q))$
still has a $q=2k_F$ divergence at a 
critical transport rate $\tau _{c}^{-1}(T)$
{\it even for non--zero single--particle relaxation rate 
$\tau _{qp}^{-1} < \tau ^{-1}$ and at finite $T$}. 
The parameter $\tau ^{-1}$ is varied experimentally by changing 
the  composition of the alloy. 

\section{Discussion and comparison with experiments}
Fourier transforming $1-1/\veps (q)$ to obtain $\rho (r)$ \cite{mahan.90} 
shows that for incomplete Fermi surface--Jones zone matching, i.e.
small $\tau ^{-1}$, the quantum corrections generate density oscillations
$\rho ^{(1)}(r) \propto - \mbox{cos}(2k_Fr)/r^{3}$, which overcompensate 
the conventional FOs, implying a phase shift of $\varphi = \pi$ 
\cite{kroha.95}. As
$\tau ^{-1} \rightarrow \tau _{c}^{-1}$, the increasing $2k_F$
peak of $1/\veps (q)$ leads in addition to density oscillations   
$\rho ^{(1)}(r) \propto \mbox{sin}(2k_Fr)/r^{2}$, so that in the
vicinity of $\tau _{c}^{-1}$
\begin{eqnarray}
\rho (r)\propto -{\mbox{cos}(2k_Fr)\over (2k_Fr)^{3}} + 
         A(\tau ^{-1}){\mbox{sin}(2k_Fr)\over (2k_Fr)^{2}},
\label{rho}
\end{eqnarray}
with $A(\tau ^{-1})\simeq 0.343\pi (1-\tau ^{-1}/\tau _{c}^{-1})^{-1/2}$.
The exponent $1/2$ is characteristic for diffusive behavior.
Thus, the FOs are shifted by $\varphi =\pi -\mbox{tan}^{-1}[2k_Fr~A(\tau ^{-1})]
\simeq \pi/2 + 1/(2k_FrA)$, i.e. the diverging
Friedel amplitude necessarily goes hand in hand with $\varphi = \pi/2$.
Note that, in contrast to the conventional FOs, this divergence 
is robust against damping due to finite $T$ or disorder.
The point where the amplitude $A$ diverges should be identified with the CAT,
since at this point the fluctuations of the Friedel potential also become
large, allowing the system to find its crystalline ground state.
This resolves in a natural way 
the problems (1)-(3) mentioned in the introduction. 

\begin{figure} 
\centerline{\psfig{figure=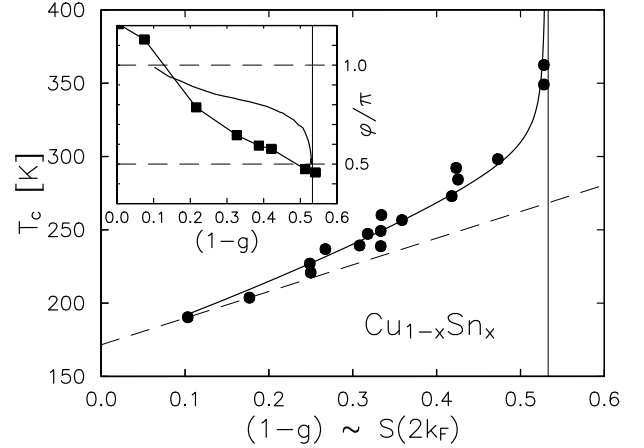,rheight=6.5cm}}
\caption{
Crystallization temperature $T_{c}$
as a function of the DOS suppression at $\veps _F$, $(1-g)$. 
Data points represent $T_{c}$ for a-$Cu_{1-x}Sn_x$ [1]. 
The solid curve is the fit of the present theory 
(see text). Vertical line: position of CAT.
The inset shows the phase shift $\varphi$ of the first maximum
of the charge density distribution $\rho (r)$.
Solid line: theory. Data points with solid line: measurements [1] for
a-$Cu_{1-x}Sn_x$.
\label{fig:Tcryst}}
\end{figure} 
For a direct comparison with experiments the control parameter of the
theory, $\tau ^{-1}$, must be translated into a parameter, 
which is experimentally accessible: 
It follows from the scattering theory \cite{kroha.90,kroha.95} that 
$\tau ^{-1} = \tau _o^{-1} + \gamma ~S(2k_F)$, where the peak of the ionic
structure factor $S(q=2k_F)$ controls the backscattering amplitude, 
$\gamma $ is a constant, and $\tau_o^{-1}$ is an offset due to 
momentum independent scattering. $S(2k_F)$ in turn is 
proportional \cite{haeuss.92} to
the measured, structure--induced suppression of the DOS $N(\veps _F)$ 
at the Fermi level, $1- N(\veps _F)/N_o(\veps _F)\equiv 1-g$, 
compared to the free electron gas, $N_o(\veps _F)$.  
The resulting fit of the crystallization temperature $T_{c}$ is shown in 
Fig. \ref{fig:Tcryst}, where the contribution to the stability coming from
the pseudogap formation is assumed to be linear in $(1-g)$ (dashed line).
The inset shows the calculated phase shift $\varphi$ 
and the measured shift of the atomic nearest neighbor position
relative to the position of the 
first conventional Friedel minimum, $a_o = \pi / k_F$.
Note that there is no adjustable parameter in $\varphi$. 
The general behavior of the shift
is well explained by the theory; however, the experimental data approaches
$\varphi = \pi/2$ faster than predicted. This might be attributed to the
fact that, as seen from the discussion after Eq. (\ref{rho}), the 
higher--order Friedel minima approach $\varphi = \pi /2$ faster than the
first one. In this light, the agreement between
theory and experiment is remarkably good.
 
The structural similarities \cite{poon.92} 
between amorphous alloys and {\it i}--quasicrystals
suggest that the quantum effect discussed above may be important 
in the latter systems as well. In fact, quasicrystals seem to 
fulfill all the necessary preconditions for this effect to
occur, i.e. effectively diffusive electron motion \cite{piechon.96}
and $\tau ^{-1} \gg \tau _{qp}^{-1}$. The latter is 
supported by the Fermi surface matching, i.e. by the experimental
observation \cite{davydov.96} and theoretical prediction 
\cite{ashcroft.87,fuji.91,hafner.92} 
of structure--induced pseudogaps. Moreover, another
more commonly known effect of disorder--enhanced Coulomb interaction, 
the $\sqrt{|E-\veps _F|}$ behavior of the DOS in the 
pseudogap \cite{altsh.79}, 
may have been already observed in {\it i}--quasicrystals by 
tunneling measurements of the DOS \cite{davydov.96}.   
It is proposed to include the enhanced Friedel potential 
calculated in the present work in the pseudopotential of more
quantitative {\it ab initio} calculations.	 

Numerous discussions with A.~G.~Aronov, P.~H\"aussler,
A.~Huck, T.~Kopp, Ch.~Lauinger, and P.~W\"olfle are gratefully 
acknowledged. This work is supported by DFG through SP Quasikristalle.  


\newpage

\end{multicols}

\end{document}